%% file: main.tex
\documentclass[11pt]{article}
\usepackage[utf8]{inputenc}
\usepackage{amsmath}
\usepackage[dvipsnames]{xcolor}
\usepackage{amsthm}
\usepackage{pdflscape}
\usepackage{pgfplots}
\pgfplotsset{compat=1.18}
\usepackage{mathrsfs}
\usepackage{tikz}
\usepackage{amssymb}
\usepackage{forest}
\usepackage[hmargin=1.25in]{geometry} 
\usepackage{graphicx}
\usepackage{float}
\usepackage[colorlinks, backref=page]{hyperref}
\usepackage{enumerate}
\usepackage{cleveref}
\usepackage[shortlabels]{enumitem}
\usepackage{algorithm}
\usepackage{algpseudocode}
\usepackage{thmtools}
\usepackage{thm-restate}
\usepackage{titlesec}
\usepackage{alphalph}
\usepackage{comment}

\usepackage{caption}
\usepackage[indention=20pt]{subcaption}

\input{macros}

\title{The Greedy Coin Change Problem}

\author{Shreya Gupta\thanks{University of California, San Diego. \href{mailto:sfgupta@ucsd.edu}{sfgupta@ucsd.edu}.},
Boyang Huang\thanks{University of California, San Diego. \href{mailto:boh002@ucsd.edu}{boh002@ucsd.edu}.},
Russell Impagliazzo\thanks{University of California, San Diego. \href{mailto:rimpagliazzo@ucsd.edu}{rimpagliazzo@ucsd.edu}. Supported by NSF Award AF: Medium 2212136}
}

\begin{document}

\maketitle
\pagenumbering{gobble}

\begin{abstract}
    \input{abstract}
\end{abstract}
\newpage

\pagenumbering{arabic}
\setcounter{page}{1}

\input{intro}

\input{prelims}

\input{P_completeness}

\input{conclusion}

\newpage
\bibliographystyle{alpha}
\bibliography{references}

\end{document}

%% file: macros.tex
\newcommand{\cC}{\mathcal{C}}

\newcommand{\cG}{\mathcal{G}}

\newcommand{\N}{\mathbb{N}}

\newcommand{\qacc}{q_{\textup{accept}}}
\newcommand{\qrej}{q_{\textup{reject}}}

\theoremstyle{plain}
\newtheorem{theorem}{Theorem}[section]

\newtheorem{proposition}[theorem]{Proposition}
\newtheorem{lemma}[theorem]{Lemma}

\theoremstyle{definition}
\newtheorem{definition}[theorem]{Definition}
\newtheorem{assumption}[theorem]{Assumption}

\theoremstyle{remark}
\newtheorem{remark}[theorem]{Remark}

\newcommand{\poly}{{\rm poly}}

\newcommand{\dtime}{{\bf DTIME}}

\newcommand{\ptime}{{\bf P}}

\newcommand{\logspace}{{\bf L}}
\newcommand{\nc}{{\bf NC}}
\newcommand{\np}{{\bf NP}}

\newcommand{\ccopy}{c_{\textup{copy}}}
\newcommand{\ctrans}{c_{\textup{transition}}}
\newcommand{\ctransl}{c_{\textup{transition}}^{\textup{left-end}}}
\newcommand{\gcc}{\mathrm{GCC}}
\newcommand{\cc}{\mathrm{CC}}

%% file: abstract.tex
The \emph{Coin Change} problem, also known as the Change-Making problem, is a well-studied combinatorial optimization problem, which involves minimizing the number of coins needed to make a specific change amount using a given set of coin denominations.
A natural and intuitive approach to this problem is the greedy algorithm.
While the greedy algorithm is not universally optimal for all sets of coin denominations, it yields optimal solutions under most real-world coin systems currently in use, making it an efficient heuristic with broad practical applicability.
Researchers have been studying ways to determine whether a given coin system guarantees optimal solutions under the greedy approach, but surprisingly little attention has been given to understanding the general computational behavior of the greedy algorithm applied to the coin change problem.

To address this gap, we introduce the \emph{Greedy Coin Change} problem and formalize its decision version: given a target amount $W$ and a set of denominations $\cC$, determine whether a specific coin is included in the greedy solution.
We prove that this problem is $\ptime$-complete under log-space reductions, which implies it is unlikely to be efficiently parallelizable or solvable in limited space.

%% file: intro.tex
\section{Introduction}
\label{sec:introduction}

The \emph{Coin Change} problem, also known as the Change-Making problem, is a classical combinatorial optimization problem that arises frequently in both theoretical computer science and real-world applications.
The problem asks for a way to make a specific target amount of change using the fewest number of coins from a given set of denominations.
Formally, given a target change amount $W \in \N$\footnote{In this paper, we use the convention that $\N = \{1, 2, \dotsc \}.$} and a set of $n$ coin denominations $\cC := \{c_1, c_2, \dotsc, c_n\} \subseteq \N$, the goal is to choose the minimum number of coins from $\cC$ that sum exactly to $W$, where each coin can be chosen arbitrarily many times.

The coin change problem is known to be weakly $\np$-hard \cite{garey1990guide-np-completeness, lueker1975two} and equivalent to the Unbounded Knapsack problem, a variant of the well-known $\np$-complete 0-1 Knapsack problem.
K\"unnemann, Paturi, and Schneider studied its fine-grained complexity, showing that the problem is sub-quadratically equivalent to $(\min, +)$-convolution \cite{künnemann2017finegrainedcomplexityonedimensionaldynamic}.
Despite its computational hardness, this problem has also served as a canonical example in algorithm design, showcasing the strengths and limitations of fundamental algorithmic paradigms such as greedy algorithms and dynamic programming.
For example, it has been widely adopted in computer science textbooks to introduce and contrast these algorithmic techniques \cite{clrs2009introalgs, goodrich2014algotihm-design-applications}.
Beyond its pedagogical value, the coin change problem is widely used in practical applications, such as distributing change in grocery stores, vending machines, and shipping systems.

Extensive research has been studying the computational complexity of finding the optimal solution to the coin change problem.
Dynamic programming is a well-known classical approach to solving the coin change problem optimally, yielding a pseudo-polynomial $O(nW)$-time algorithm \cite{wright1975coin-making-dp}.
More recently, Chan and He proposed a deterministic algorithm with $O(W \log W \log \log W)$ running time as well as a randomized algorithm with $O(W \log W)$ expected running time with an application of convolution (i.e., Fast Fourier Transform) \cite{chan2020onthechangemaking}.
Later, they extended their results to address both the original single-target version of the problem and the more general all-targets version, along with improved algorithms \cite{chan2021moreonthechangemaking}.

In contrast, the greedy algorithm is a natural and intuitive approach to the coin change problem \cite{chang1970alg}, despite not always guaranteeing an optimal solution.
Another line of research has been focusing on the conditions under which the greedy algorithm yields an optimal solution.
Chang and Gill identified a range of target amounts $W$ for a given set of coin denominations where counterexamples to the optimality of the greedy algorithm can occur and proposed a procedure to test for such cases \cite{chang1970alg}.
This was later improved by Kozen and Zaks, who established tight bounds for counterexample ranges and gave a simpler testing procedure \cite{kozen1994optimal}.
Additionally, Person proposed a polynomial-time $O(n^3)$ algorithm to determine whether the greedy algorithm is optimal given a coin system \cite{person2005polydetermineoptimal}.
Still, the greedy algorithm is optimal for nearly all major real-world currency systems, making it an efficient heuristic with broad practical applicability.

However, although much has been studied about the \emph{optimality} of the greedy algorithm, there has been surprisingly limited research into the \emph{computational complexity} of \emph{simulating} the greedy algorithm.
By \emph{simulation}, we refer to any algorithm that produces the same set of coin selections as the greedy approach.
Moreover, it is worth noting the distinction between studying the computational complexity of the coin change \emph{problem} itself versus studying the complexity of simulating the greedy \emph{algorithm} applied to this problem.
Given the sequential nature of greedy algorithms, where each step makes a locally optimal choice by selecting the largest coin not exceeding the remaining target amount, a natural question to consider is whether the greedy algorithm on the coin change problem is efficiently parallelizable.

To formalize this question, we introduce the \emph{Greedy Coin Change} problem, which asks to output the set of coins selected by the greedy strategy.
We define its decision version to take a special query coin as an additional input and ask whether this special coin is selected by the greedy algorithm.
It turns out that the decision version of the greedy coin change problem is indeed $\ptime$-complete under log-space reduction, which confirms the intuition about the inherently sequential nature of the greedy algorithm.

\begin{restatable}[$\ptime$-completeness of the Greedy Coin Change Problem]{theorem}{GCCPC}
    \label{thm:gcc-p-complete}
    The Greedy Coin Change problem is P-complete under log-space reduction.
\end{restatable}

The study of $\ptime$-completeness problems emerged from an interest in understanding the limits of parallel computation \cite{greenlaw1995limits, cook1985taxonomy}.
While problems in the class $\ptime$ are solvable in polynomial time by a deterministic Turing machine, not all such problems are known to be efficiently parallelizable.
The greedy coin change problem shares similarities with other well-known $\ptime$-complete problems, such as the Lexicographically First Maximal Independent Set (LFMIS) problem \cite{cook1985taxonomy, uehara1997measure, uehara1999another, uehara1999measure, miyano1989list}, in that they all involve a set of objects of greedy nature within a combinatorial optimization problem.

\paragraph*{Organization}
In \Cref{sec:prelims}, we formally define the greedy coin change problem as well as log-space reductions.
In \Cref{sec:p-completeness}, we prove the $\ptime$-completeness of the greedy coin change problem.

%% file: prelims.tex
\section{Preliminaries}
\label{sec:prelims}

In this section, we provide the formal definitions for the concepts relevant to this paper.

\subsection{Greedy Coin Change Problem}

The classical Coin Change problem asks to determine the minimum number of coins from a given set of denominations that sum to a specified target change amount.

\begin{definition}[Coin Change $\cc(W, \cC)$]\label{def:cc}
    Given a target change amount $W \in \N$ and a set of $n$ coin denominations $\cC = \{c_1, c_2, \dotsc, c_n\} \subseteq \N$, output the optimal combination of coins that sum up to $W$.
    Formally, the goal is to compute non-negative integers $x_1, x_2, \dotsc, x_n$ to
    \begin{align*}
    &\text{minimize} \quad \sum_{i=1}^n x_i, \\
    &\text{subject to} \quad \sum_{i=1}^n x_ic_i = W.
    \end{align*}
    Without loss of generality, we assume that a coin of value 1 is always in $\cC$, so that there always exists a combination of coins that sum up to $W$ for all $W \in \N$.
\end{definition}

Instead of the optimal combination of coins, we study the property of the \emph{greedy set of coin change} in this paper.

\begin{definition}[Greedy Set of Coin Change]\label{def:greedy-set}
    Given a target change amount $W \in \N$ and a set of $n$ coin denominations $\cC = \{c_1, c_2, \dotsc, c_n\} \subseteq \N$, the \emph{greedy set of coin change} $\cG$ is constructed recursively as follows, starting with $\cG =  \emptyset$: for any remaining change amount $W' \leq W$, add to $\cG$ the largest coin $c_i \in \cC$ subject to $c_i \leq W'$ and set $W' \gets W' - c_i$.
\end{definition}

We are now ready to define the Greedy Coin Change problem.

\begin{definition}[Greedy Coin Change Problem $\gcc(W, \cC, c^*)$]\label{def:gcc}
    Given a target change amount $W \in \N$ and a set of $n$ coin denominations $\cC = \{c_1, c_2, \dotsc, c_n\} \subseteq \N$, output the greedy set of coin change $\cG$ with respect to $W$ and $\cC$.

    The \emph{decision} version of the Greedy Coin Change problem takes a special coin $c^* \in \cC$ as an additional input and asks whether $c^* \in \cG$.
\end{definition}

Unless otherwise noted, we work with the decision version of the greedy coin change problem in the rest of this paper.

\subsection{P-completeness and Log-space Reductions}

Formally, a decision problem $L$ is said to be $\ptime$-complete if it belongs to the complexity class $\ptime$ and every problem in $\ptime$ reduces to $L$ under some suitable notion of reductions.
Generally, the reductions considered for $\ptime$-completeness are log-space reductions and $\nc$-reductions.
We first provide a formal definition for log-space reductions, which is the notion of reduction we use for showing the $\ptime$-completeness of the greedy coin change problem.

\begin{definition}[Log-space Reductions \cite{sipser1996introduction}]
A \emph{log-space transducer} is a Turing machine with a read-only input tape, a write-only, write-once output tape, and a read/write work tape.
The head on the output tape cannot move leftward, so it cannot read or overwrite what it has written.
The work tape may contain $O(\log n)$ symbols.

A log space transducer $M$ computes a function $f: \Sigma^* \to \Sigma^*$, where $f(W)$ is the string remaining on the output tape after $M$ halts when it is started with $w$ on its input tape.
We call $f$ a \emph{log-space computable function}.

Language $A$ is \emph{log-space reducible} to language $B$, written $A \leq_\logspace B$, if $A$ is mapping reducible to $B$ by means of a log-space computable function $f$.
\end{definition}

On the other hand, the complexity class $\nc$ is defined with respect to parallel computation.
Roughly speaking, a decision problem is in $\nc$ if it is decidable in poly-logarithmic time on a parallel computer with a polynomial number of processors.
Similarly, an $\nc$-reduction is one that is computable in poly-logarithmic time on a parallel computer with a polynomial number of processors.
While a $\ptime$-completeness result under $\nc$-reductions for the greedy coin change problem may seem more directly relevant to our initial question of whether there is an efficient parallelization of the greedy algorithm on the coin change problem, it is known that log-space reductions are \emph{stronger} than $\nc$-reductions, in the following sense:

\begin{theorem}[\cite{sipser1996introduction}]
\label{thm:log-space-reduction-stronger}
    Let $A, B$ be two languages.
    If $A \leq_\logspace B$ and $B \in \nc$, then $A \in \nc$.
\end{theorem}

\begin{remark}
    \Cref{thm:log-space-reduction-stronger} also holds if we replace $\nc$ with $\logspace$ \cite{sipser1996introduction}.
\end{remark}

In particular, \Cref{thm:log-space-reduction-stronger}, combined with our result, means that if the greedy algorithm on the coin change problem is efficiently parallelizable (i.e., in $\nc$), then all problems decidable in polynomial-time are also in $\nc$; however, under the unproven yet widely believed assumption that $\nc \subsetneq \ptime$, this would imply that the greedy algorithm on the coin change problem is not efficiently parallelizable.

%% file: P_completeness.tex
\section{P-completeness of the Greedy Coin Change Problem}\label{sec:p-completeness}

Recall our main result on the hardness of the greedy coin change problem:

\GCCPC*

Our proof uses a generic reduction from a polynomial-time Turing machine.
The high-level idea is to construct an instance of the greedy coin change problem such that the remaining change amounts represent configurations and coins represent transitions of the machine.
Let us first formalize the Turing machine model we use in the proof and state some necessary assumptions.

\subsection{Turing Machine}\label{sec:turing-machine}
In this reduction, we use the standard deterministic single-tape Turing machine model.
A Turing machine is given as \(M = \langle Q, \Gamma , \Sigma , \delta , q_0, \qacc, \qrej \rangle \), where \(Q\) is the set of states, \(\Gamma\) is the set of tape alphabets, \(\Sigma \subseteq \Gamma  \setminus  \{\perp\}\) is the set of input alphabets, $\perp \in \Gamma$ is the blank symbol, \(\delta : Q \setminus F \times \Gamma  \to Q \times \Gamma  \times \{L, R\}\footnote{$L$ indicates that the tape head moves left by one tape cell and $R$ indicates right.}\) is the transition function, \(q_0 \in Q\) is the start state, \(\qacc \in Q\) is the accept state, and \(\qrej \in Q\) is the reject state.
We use \(F := \{\qacc, \qrej\} \subseteq Q\) to denote the set of \emph{halting states}.
Moreover, we make the following assumptions about $M$.
\begin{assumption}\label{assumption:tm-left-endmarker}
    We assume that there is always a left endmarker $\$$ in the first tape cell, and the tape head always moves right whenever it reads $\$$.
\end{assumption}
\noindent This ensures that the tape head always moves in the direction indicated by the transition function. (In the standard Turing machine model, the tape head could get stuck in the same tape cell if it tries to move left while over the leftmost tape cell.) 
\begin{assumption}\label{assumption:tm-tapehead-move-to-the-front-when-terminating}
    We assume that when \(M\) decides to transition into a halting state, the tape head first moves all the way to the left endmarker, then moves right and overwrites the symbols in the second and third tape cells with $\perp$, and finally stops over the second tape cell in the corresponding halting state.
\end{assumption}
\noindent We note that both assumptions would only slow down the running time by a constant factor.

\subsection{Greedy Coin Change Instance Construction}\label{sec:gcc-instance-construction}

In this section, we formally construct the greedy coin change instance $\gcc(W, \cC, c^*)$ used in the reduction.
The goal is to construct $W$ and $\cC$ so that the greedy set of coins $\cG$ with respect to $W$ and $\cC$ exactly represent the transitions of a Turing machine, hence simulating its computation.

Let \(L\) be a language in $\ptime$ and say that $L \in \dtime(n^\ell)$ for some $\ell$. 
Let $M$ be the Turing machine that decides $L$ in $O(n^{\ell})$ time on inputs of size $n$.
For this construction to work, we need the \emph{exact} computation time (upper bound) of $M$ instead of the asymptotic one, so let $C_M > 0$ be the constant such that \(M\) halts within \(C_{M} \cdot n^{\ell}\) time for sufficiently large input size $n$.
Now, let $x = (x_1, \dotsc, x_n)$ be an arbitrary input of size $n$ and let $T = C_{M} \cdot n^{\ell}.$

\paragraph*{Bit-string Representation of Configuration}
A \emph{configuration} $C$ of the Turing machine $M$ at any given time contains the following information: 1) the current state $q \in Q$, 2) the current tape head position $i \geq 1$, and 3) the current tape content.
We say that $C$ is a \emph{halting configuration} if $q \in F$ is a halting state.
Since the Turing machine $M$ halts within time $T$, it uses at most $T$ tape cells throughout the computation.
Therefore, in any configuration $C$ of $M$, the tape head position satisfies $1 \leq i \leq T$, and the tape content can be represented as a $T$-bit string of symbols $a_1 a_2 \dotsb a_T \in \Gamma^T$.
A typical way of representing $C$ is
\[
    C = a_1 \dots a_{i-1} (q a_i) a_{i+1} \dots a_T.
\]
To represent configurations using bit-strings, we first identify the set of states and the set of alphabets separately with numbers
\[
    Q \overset{f}{\cong} \{1, \dotsc, s\} \quad \text{and} \quad \Gamma \overset{g}{\cong} \{1, \dotsc, k\}
\]
where $s := |Q|$ and $k := |\Gamma|$. Here, $f$ and $g$ can be any bijections that map the elements of $Q$ and $\Gamma$ to the numbers $\{1, \dotsc, s\}$ and $\{1, \dotsc, k\}$, respectively.
Next, we identify the set of all state-alphabet pairs with
\begin{align*}
    Q \times \Gamma &\cong \{k+1, \dots, 2k, \dots, (s + 1)k\}\\
    (q, a)  &\mapsto f(q)\cdot k + g(a)
\end{align*}
Now, if we choose a large enough base $B \geq (s+1)k + 2$, then we can view the configuration $ C = a_1 \dots a_{i-1} (q a_i) a_{i+1} \dots a_T$ as a $T$-bit string in base $B$.
The next two propositions follow directly from the definitions but will turn out to be very useful in the proof later.

\begin{proposition}\label{prop:pair-larger-than-alph}
    Any state-alphabet pair $qa$ is mapped to a strictly larger value than any single alphabet $a'$.
\end{proposition}

\begin{proposition}\label{prop:B-1-large}
    Any state $q$, alphabet $a$, and state-alphabet pair $qa$ is mapped to a value strictly less than $B - 1$.
\end{proposition}

Having defined the bit-string representations of the configurations, we are now ready to define the initial change amount $W$ and the coin set $\cC$ of the greedy coin change instance.
Recall that intuitively we want each remaining change amount (including the initial amount) to represent a configuration and each coin to represent a valid transition between configurations, so that in a change-making process, subtracting the value of a coin from some remaining change amount is equivalent to transitioning from one configuration to another.

\paragraph*{Defining the Initial Change Amount}
To differentiate between configurations of different time steps, we lift these bits by left-shifting $(T-j)T$ bits for time step $j$:
\[
 a_1 \dots a_{i-1} (q a_i) a_{i+1} \dots a_T \cdot B^{(T-j)T}.
\]
The reversed shifting factor $T - j$ for time $j$ (instead of a factor of $j$) is due to the fact that the remaining change amount keeps decreasing during a change-making process, so we want the higher bits to represent the configurations earlier in the computation.
This naturally leads to the following definition of the initial change amount $W$: 
\begin{equation}\label{eqn:initial-change-amount}
    W :=  (q_0 \$) x_1 \cdots x_n \underbrace{\perp \cdots \perp}_{T - n - 1 \text{ times}} \cdot B^{(T - 1)T},
\end{equation}
where we recall that $q_0$ is the initial state of $M$, \$ is the left endmarker, $x = (x_1, x_2, \dotsc, x_n)$ is the input to $M$, and $\perp$ is the blank symbol.

\paragraph*{Defining the Coin Set}

Recall that we want to define coins to represent valid transitions between configurations.
For any configuration $C$ of the Turing machine $M$, we use $\Delta(C)$ to denote the unique next configuration of $C$ according to the (deterministic) transition function $\delta$ of $M$.
If $C$ is a halting configuration, we let $\Delta(C) = C$.

If a remaining change amount corresponds to $C$ at some time step $j$, represented by some range of $T$ bits, then we want to define a coin that subtracts the current configuration $C$ from the corresponding range and add the new configuration $\Delta(C)$ to a lower range of $T$ bits:
\[
C \cdot B^{(T-j)T} \xrightarrow[\text{single coin change}]{C \cdot B^{(T-j)T} - \Delta(C) \cdot B^{(T-j - 1)T}} \Delta(C) \cdot B^{(T-j - 1)T}.
\]
Naively, defining the coin set under this approach requires at least one coin for each different configuration $C$.
However, since all we know about $M$ is that it runs in polynomial time, thus $\poly(n)$ is the best space complexity upper bound we could assume for $M$.
But this means that there are exponentially many possible configurations of $M$, which are too many to output with a log-space transducer.
In order to output the coin set with a log-space transducer, we need some sort of uniformity in the coin set.
To do this, we leverage the locality of Turing machine transitions and break down each transition into a sequence of roughly $T$ \emph{local} transitions, which we explain in detail below.

Consider an arbitrary non-halting configuration $C$ of the Turing machine $M$ and the transition from $C$ to $\Delta(C)$.
Say that the tape head is at location $h$ in the configuration $C$.
Note that for most of the tape cells that are not in the neighborhood of $h$\footnote{We say that a tape cell at position $i$ is in the neighborhood of the tape head location $h$ if $|i - h| \leq 1$.}, we simply need to copy over the same alphabet symbol to the next lower range of bits.
This is because the tape head cannot move to location $i$ in a single transition under our standard model of Turing machine (see \Cref{sec:turing-machine}).
On the other hand, for the tape cells $i$ that are within the neighborhood of $h$, the tape head could either move into or away from location $i$.
To correctly simulate this local transition, we define a coin that updates the consecutive three tape cells of the neighborhood of $h$ all at once. 
Concretely, the coin set $\cC$ contains precisely the following two types of coins:
\begin{itemize}
    \item \textbf{Copy coin.} For each $1 \leq i, j \leq T$ and $a \in \Gamma$, define a copy coin
    \begin{equation}\label{eqn:copy-coin}
        \ccopy(a, i, j) := 0^{i-1} a  0^{T-i} \cdot B^{(T-j)T}
         - 0^{i-1} a  0^{T-i} \cdot B^{(T-j-1)T}.
    \end{equation}
    
    \item \textbf{Transition coin.} For each $2 \leq i \leq T-1$, $1 \leq j \leq T$, $q \in Q$, and $a^-, a, a^+ \in \Gamma$, define a transition coin
    \begin{equation}\label{eqn:trans-coin}
        \ctrans(q, a^-, a, a^+, i, j) := 0^{i-2}a^-(q a)a^+ 0^{T-i-1} \cdot B^{(T-j)T} - s \cdot B^{(T-j-1)T},
    \end{equation}
    where $s$ is the update to the lower range of bits defined as
    \[  
    s :=\begin{cases}
            0^{i-2}(q' a^-)a'a^+ 0^{T-i-1} & \text{ if } q \in Q \setminus F \text{ and }\delta(q, a) = (q', a', L)  \\
            0^{i-2}a^-a'(q' a^+) 0^{T-i-1} & \text{ if } q \in Q \setminus F \text{ and }\delta(q, a) = (q', a', R) \\
            0^{i-2}a^-(q a) a^+ 0^{T-i-1} & \text{ if } q \in F
        \end{cases}.
    \]
    In addition, we need a special transition coin for the left end of the tape.
    Due to \Cref{assumption:tm-left-endmarker}, the leftmost tape cell always contains the special symbol $\$$, and the tape head always moves right whenever it is over $\$$. 
    Thus, we only need to define for $q \in Q \setminus F$, $a^+ \in \Gamma$, and $1 \leq j \leq T$, the following coin
    \begin{equation}\label{eqn:end-trans-coin}
        \ctransl(q, a^+, j) := (q \$) a^+ 0^{T-2} \cdot B^{(T-j)T} -   \$ (q'a^+) 0^{T-2} \cdot B^{(T-j - 1)T},
    \end{equation}
    where $(q', \$, R) = \delta(q, \$)$ (recall \Cref{assumption:tm-left-endmarker}).
\end{itemize}

\begin{remark}
    For convenience, we abuse the notation when $j = T$ and consider $B^{(-1)T}$ as 0.
    Also for clarity, we are padding with zeros at the front to align with our bit-string representation of configurations, although they are not necessary for the arithmetic.
\end{remark}

Note that we do not need a special transition coin for the right end due to \Cref{assumption:tm-tapehead-move-to-the-front-when-terminating}: if the tape head needs to move back to the second cell before the machine terminates within time $T$, it is impossible for the tape head to ever move over the $T$-th cell.

\paragraph*{Defining the Special Query Coin}
Finally, we define the special query coin as the following transition coin:
\begin{equation}\label{eqn:query-coin}
    c^* := \ctrans(\qacc, \$, \perp, \perp, 2, T) = \$(\qacc \perp) \perp \underbrace{0 \cdots 0}_{T - 3\textup{ times}}.
\end{equation}

This completes the definition of the greedy coin change instance $\gcc(W, \cC, c^*)$.
Next, we proceed to prove the correctness of the reduction.

\subsection{Proof of Correctness}
The correctness of the greedy coin change instance $\gcc(W, \cC, c^*)$ constructed in the previous section can be stated as the following lemma.

\begin{lemma}\label{lem:gcc-sim-TM-acceptance}
    The Turing machine $M$ accepts on input $x$ if and only if $c^*$ belongs to the greedy set of coins $\cG$ with respect to $W$ and $\cC$.
\end{lemma}

To prove \Cref{lem:gcc-sim-TM-acceptance}, it suffices to show that the greedy coin change instance $\gcc(W, \cC, c^*)$ can simulate any valid transitions between configurations.
Then, the result follows from a simple inductive argument.

\begin{lemma}\label{lem:gcc-sim-transition}
    Let $C$ be a configuration of $M$ at some time step $j$.
    Then, with respect to the remaining change amount
    \[
        W_C := C \cdot B^{(T-j)T},
    \]
    which represents $C$ at time $j$, and the coin set $\cC$, the greedy coin set $\cG$ simulates the transition from $C$ to $\Delta(C)$ through a sequence of $T-1$ or $T-2$ coin changes.
    Formally, this means that the next $T-1$ or $T-2$ coins to be included in the greedy coin set $\cG$ reduces the remaining change amount to $W_{\Delta(C)} := \Delta(C) \cdot B^{(T - j - 1)T}$, which represents $\Delta(C)$ at the next time step.
\end{lemma}
\begin{proof}
    There are four different forms of intermediate change amount $W' \leq W_C$ during a transition from $W_C$.
    In each case, we show that the greedy set of coins $\cG$ includes the desired coin.
    The result then follows by a simple induction and combining multiple cases to cover the entire transition process from $W_C$ to $W_{\Delta(C)}$.

    Before proceeding to the main proof, we note that each coin value for $1 \leq j < T$ is defined as a difference of two values in \Cref{sec:gcc-instance-construction} to match the idea of local transitions; that is, each coin subtracts some bits from a higher range and then adds some (other) bits to a lower range.
    In this proof, it might be more helpful to consider the actual values of the coins rather than viewing them as differences.
    This is because showing that the desired coin is included in $\cG$ amounts to showing that it is the largest coin in the coin set $\cC$ that does not exceed the remaining amount $W'$, which involves comparing the magnitudes of bit-strings.
    To illustrate, we use copy coins as an example. For $1 \leq j < T$, the copy coin $\ccopy(a, i, j)$ can be equivalently written as follows (see \Cref{eqn:copy-coin} for the original definition):
    \[
        \ccopy(a, i, j) = (a-1) \underbrace{(B-1) \cdots (B-1)}_{(T-1) \textup{ times}} (B-a) \underbrace{0 \cdots 0}_{(T-i) \textup{ times}} \cdot B^{(T-j-1)T}.
    \]
    The key observations are:
    \begin{itemize}
        \item For copy coins, the leading bit is $a-1$; for transitions coins, the leading three bits are $a^-(qa)(a^+ - 1)$; for left-end transition coins, the leading two bits are $(q\$)(a^+ - 1)$.
        \item In all cases, the above leading bits are immediately followed by a block of $(B-1)$-bits.
    \end{itemize}
    We now proceed to analyze the four cases.
    \begin{enumerate}
        \item $W' = (q\$) a_2 a_3 \cdots a_T \cdot B^{(T-j)T}$.
        This represents the initial stage where $W' = W_C$ and no coins have been chosen yet.
        Consider the candidate coin $c_0 := \ctransl(q, a_2, j) \leq W'$, whose leading three bits are $(q\$)(a_2-1)(B-1)$.
        We show that there is no $c \in \cC$ that can beat $c_0$ under $W'$, i.e. satisfying $c_0 < c \leq W'$.
        
        If $c$ is a copy coin or a transition coin, then its leading bit is bounded above by the value of an alphabet, which is strictly less than the leading bit $(q\$)$ of $c_0$ (\Cref{prop:pair-larger-than-alph}).
        On the other hand, if $c$ is some other left-end transition coin, then its leading three bits must be at least $(q\$) a_2 (B-1)$ in order to beat $c_0$.
        But then $B - 1 > a_3$ due to \Cref{prop:B-1-large}, and so $c > W'$.
        Therefore, $c_0$ is included in $\cG$, as desired.

        \item $W' = 0^{i-1}a_i\cdots a_T \cdot B^{(T-j)T} + s \cdot B^{(T-j-1)T}$ for some partial update $s \in B^T$.
        This happens when the bit at the tape head location has already been moved to a lower range.
        We consider the candidate coin $c_0 := \ccopy(a_i, i, j) \leq W'$, whose leading two bits are $(a_i - 1) (B-1)$.
        
        Suppose for contradiction that there exists a coin $c \in \cC$ that beats $c_0$ under $W'$.
        It is not hard to see that all the left-end transition coins are either too large or too small as their leading bits are not at the correct positions.
        If $c$ is a transition coin, then its leading bit must either be $a_i$ or $a_i - 1$ to beat $c_0$.
        If the leading bit of $c$ is $a_i$, then its next bit is a state-alphabet pair which is strictly larger than $a_{i+1}$ (\Cref{prop:pair-larger-than-alph}), so $c > W'$; if the leading bit is $a_i -1$, then its next bit is strictly smaller than $B-1$ (\Cref{prop:B-1-large}), so $c < c_0$.
        Finally, if $c$ is a copy coin, then again no such coin $c$ exists due to \Cref{prop:B-1-large}.
        Therefore, the desired coin $c_0$ is included in $\cG$.

        \item $W' = 0^{i-1} a_i \cdots (q a_h) \cdots a_T B^{(T-j)T} + s \cdot B^{(T-j-1)T}$ for some partial update $s \in B^T$ and $h \geq i+2$.
        This happens when the bit at the tape head location has not been moved to a lower range yet \emph{and} the next coin should \emph{not} move the tape head bit either (since $a_i$ is not a neighbor of $a_h$).
        The exact same reasoning as in case (2) shows that $c_0 := \ccopy(a_i, i, j)$ is the coin to be included in $\cG$.

        \item $W' = 0^{i-2} a_{i-1} (qa_i) a_{i+1} \cdots a_T B^{(T-j)T} + s \cdot B^{(T-j-1)T}$ for some partial update $s \in B^T$.
        This happens when the tape head bit has not been moved to a lower range yet \emph{but} the next coin \emph{should} move the tape head bit (since $a_{i-1}$ is now neighbor to the tape head bit $(qa_i)$).
        The candidate coin for this case is $c_0 := \ctrans(q, a_{i-1}, a_i, a_{i+1}, i, j) \leq W'$, whose leading four bits are $a_{i-1} (qa_i) (a_{i+1}-1)(B-1)$.
        
        In order to beat $c_0$ under $W'$, a copy coin needs to have $a_{i-1}$ as its leading bit, but then the next bit $B-1 > (qa_i)$; a transition coin must have its leading four bits to be at least $a_{i-1}(qa_i)a_{i+1}(B-1)$, but still $B-1 > a_{i+2}$ (\Cref{prop:B-1-large}).
        On the other hand, none of the left-end transition coins have their leading bits at the correct positions unless $i = 2$.
        But even if $i = 2$, the leading bit of a left-end transition coin is $(q\$)$, which is larger than $a_{i-1} = a_1$ due to \Cref{prop:pair-larger-than-alph}.
        Therefore, the desired coin $c_0$ is included in $\cG$.
    \end{enumerate}

    Now, notice that the tape head in the configuration $C$ is either at a location $h = 1$ or some $2 \leq h \leq T - 1$.
    In the former case, we apply case (1) followed by a sequence of case (2)'s. 
    In the latter case, we apply a sequence of case (3)'s, followed by a single application of case (4), and finally a sequence of case (2)'s.
    It is not hard to verify that the sequence of coins reduces $W_C$ to the desired $W_{\Delta(C)}$ in both cases.

    The case where $j = T$ is easier as there are no negative terms in the coin values, so the same argument works by exactly matching the leading bits.
\end{proof}

Therefore, \Cref{lem:gcc-sim-TM-acceptance} follows from applying \Cref{lem:gcc-sim-transition} inductively from the initial target change amount $W$ defined in \Cref{eqn:initial-change-amount} and noting that the special query coin $c^*$ matches the configuration of the first three tape cells of $M$ at time $T$ if and only if $M$ accepts on input $x$ (\Cref{assumption:tm-tapehead-move-to-the-front-when-terminating}).
We conclude the section by proving the main result, \Cref{thm:gcc-p-complete}.

\begin{proof}[Proof of \Cref{thm:gcc-p-complete}]
    Let $L \in \dtime(n^{\ell})$ be a language in $\ptime$ and let $x$ be an input of size $n$.
    Construct a greedy coin change instance $\gcc(W, \cC, c^*)$ as defined in \Cref{sec:gcc-instance-construction}.
    The correctness follows from \Cref{lem:gcc-sim-TM-acceptance}.
    It remains to show that this instance can be outputted using a log-space transducer.

    All of our previous discussions are done in base $B$ for convenience and the intuition.
    To formally consider the work tape space complexity of the reduction, we need to transform all values into base 2.
    Note that this does not affect the correctness proof as all the arithmetic remains the same even if we change the base.
    The only thing affected is the number of bits in all the values, which may affect complexity. 
    Recall that we chose a large enough base $B$ such that $B \geq (s+1)k + 2$, where $s = |Q|$ and $k = |\Gamma|$. 
    But since both $s$ and $k$ are constant in $n$, then so is $B$.
    In fact, we may choose a large enough integer $m$ such that $B := 2^m \geq (s+1)k+2$ while still having $B = O(1)$ (and $m = O(1)$). 
    In this way, since $B$ is a power of 2, each value can be converted to binary by simply converting each base $B$ bit to $m$ binary bits, and the asymptotic number of bits remains the same.
    Moreover, each base $B$ bit is isolated into a different chunk of binary bits of length $O(1)$, so we may still view each block of $m$ binary bits as ``one single base $B$ bit.''

    Let us first consider the initial change amount defined in \Cref{eqn:initial-change-amount}
    \[
    W =  (q_0 \$) x_1 \cdots x_n \underbrace{\perp \cdots \perp}_{T - n - 1 \text{ times}} \cdot B^{(T - 1)T}.
    \]
    Clearly, $(q_0 \$)$ can be written to the output tape with $O(1)$ work tape space.
    Next, note that the input bits $x_1, \dotsc, x_n$ can be simply copied to the output tape (with some states defined to convert them into the corresponding binary values), so they do not need to appear on the work tape (which would otherwise take $O(n)$ space).
    For the remaining large chunk of $\perp$ symbols and zeros, we only need to maintain a counter and keep outputting the same value until the counter reaches a specific number.
    Note that both $T - n - 1$ and $(T-1)T$ are $\poly(n)$, so we only need a $O(\log n)$-bit counter to keep track of the number of bits outputted so far.
    Using a similar argument, we can see that the special query coin
    \[
    c^* = \ctrans(\qacc, \$, \perp, \perp, 2, T) = \$(\qacc \perp) \perp \underbrace{0 \cdots 0}_{T - 3\textup{ times}}
    \]
    can also be outputted with log-space work tape. 
    It now remains to output the massive coin set $\cC$.\

    We claim that in order to output all the coins in $\cC$ with log-space work tape, it suffices to 1) provide a uniform naming scheme for each coin that only takes $O(\log n)$ bits to keep track of, and 2) show that each coin can be outputted with log-space work tape given its name.
    Indeed, if this is the case, then we can keep track of and loop over all the names and output the corresponding coin, all done in $O(\log n)$ space.

    For 1), we have actually already provided a suitable naming scheme for the coins during definition (see \Cref{eqn:copy-coin}, \Cref{eqn:trans-coin}, and \Cref{eqn:end-trans-coin}), where each coin can be specified only using the arguments in the parentheses.
    Any component of the name that is a state $q \in Q$ or an alphabet $a \in \Gamma$ only takes $O(1)$ space to keep track of.
    The only components that have sizes depending on $n$ are the index $i$ and time step $j$.
    But both of them are at most $T = \poly(n)$ and only require $O(\log n)$ bits to record,

    For 2), we use copy coins as an example to show that $O(\log n)$ work tape space is sufficient to output the coin given its name. The argument for the other type of coins (i.e., transition coins) is similar.
    Consider a copy coin written in its actual value:
    \[
        \ccopy(a, i, j) = \underbrace{0^{i-1} (a-1)  (B-1)^{T-i}}_{\textup{first $T$ bits}} \underbrace{(B-1)^{i-1} (B-a) 0^{T-i}}_{\textup{next $T$ bits}} \cdot B^{(T-j-1)T}.
    \]
    Note that the actual value of the copy coin also contains large blocks of repeating values, with a constant number of alternations between the blocks.
    Thus, again with the same argument for outputting the initial change amount $W$, we can output the copy coin with log-space work tape.

    This completes the proof of showing that the greedy coin change problem is $\ptime$-complete under log-space reductions.
\end{proof}

%% file: conclusion.tex
\section{Conclusion and Future Work}

In this work, we have studied the computational complexity of the greedy coin change problem, adding it to the list of $\ptime$-complete problems under log-space reductions.
While the greedy algorithm serves as an efficient heuristic to the coin change problem in practical applications, our result reinforces the intuition about its inherently sequential nature.

A promising direction for future research is to explore whether \emph{succinct input representations} exist for the greedy coin change problem, particularly for the coin set.
The existence of such succinct representations could potentially lead to either better algorithms that take advantage of the succinct representation, or more efficient reductions that provide new insights into the computational complexity of the same problem when given succinctly represented inputs.

For example, \cite{dalirrooyfard2020factor} introduced a general framework of ``factored problems'' on bit strings, demonstrating wide implications for fine-grained and average case hardness of various other computational problems.
More recently, \cite{gupta2024computationalcomplexityfactoredgraphs} studied ``factored graphs'' and showed connections to parameterized complexity.
In particular, they showed that a parameterized version of the Lexicographically First Maximal Independent Set (LFMIS) problem given factored graph inputs becomes $\mathbf{XP}$-complete.
As the LFMIS problem and the greedy coin change problem are both $\ptime$-complete problems sharing the same kind of greedy nature in their definitions, do similar results about succinct representations hold for the greedy coin change problem?

In the context of the greedy coin change problem, a potential approach to succinctly representing the coin set could be arranging the bit-string representations of the coins into rows of a matrix.
Succinct representations for matrices have been studied primarily using matrix tensor products, as well as standard matrix products and summations \cite{damm2002tensor-calculus, beaudry2001tensor-circuit-eval, lohrey2017succinct-matrix-vectors}. 
The question here is whether such a matrix representation of the coin set can be efficiently factorized into smaller matrices following the succinct representations for matrices.